\documentclass[english,aps,preprint]{revtex4}%
\usepackage{amsmath}
\usepackage{amsfonts}
\usepackage{graphicx}
\usepackage{MnSymbol}

\begin{document}
\title{Evolution of strictly localized states in non-interacting quantum field theories with
	background fields}
\author{M.\ Alkhateeb and A. Matzkin}

\affiliation{Laboratoire de Physique Th\'eorique et Mod\'elisation, CNRS Unit\'e 8089, CY Cergy
	Paris Universit\'e, 95302 Cergy-Pontoise cedex, France} 

\begin{abstract}
We investigate the construction of spin-1/2 fermionic and spin-0 bosonic wave-packets
having compact spatial support in the framework of a computational quantum
field theory (QFT) scheme offering space-time solutions of the relativistic
wave equations in background fields. In order to construct perfectly localized
wave-packets, we introduce a spatial density operator accounting for particles
of both positive and negative charge.\ We examine properties of the vacuum and 
single-particle expectation values 
of this operator and compare  them to the standard QFT particle and anti-particle 
spatial densities. The formalism is illustrated by computing numerically the
Klein tunneling dynamics of strictly localized wave-packets impinging on a supercritical
electrostatic step. The density operator introduced here could be useful to
model situations in which it is desirable to avoid dealing with the infinite
spatial tails intrinsic to pure particle or anti-particle wave-packets.

\end{abstract}

\maketitle
\newpage

\section{Introduction}

Quantum field theory with a background field has been employed in various
studies, beginning with the Schwinger-Sauter effect for pair creation \cite{schw},
the excitation of vacuum by intense laser fields \cite{review}, and
the analysis of scattering problems \cite{gitman,grobe-review,grobe_bo,hansen,nakazato}.
In particular, a time-dependent non-perturbative formalism of quantum field
theory with a background potential has been developed in order to compute
numerically the dynamics of pair creation \cite{grobe-review}. This formalism
has successfully computed the rates of pair creation for spin-0 bosons and
spin-1/2 fermions resulting from the excitation of the vacuum by arbitrary
electrostatic potentials \cite{grobe_rates,grobe_arb,lv,wang}. Recently, we
have used this formalism to provide insights into the dynamics of Klein
tunneling for both fermions and bosons \cite{ourqft}.

One of the main advantages of using this formalism is the ability to account
for the time-dependent dynamics of wave packets propagating in arbitrary
electromagnetic potentials. Typically, one considers an approximately
spatially localized wave-packet of unit charge representing the initial
particle (say an electron) scattering on a supercritical potential. The
corresponding (e.g. fermionic) field operator is then used to construct charge
density operators that give the evolution of the fermionic or anti-fermionic
states as the wave-packet scatters on the potential producing
particle/anti-particle pairs.

It is well-known that a wave packet composed solely of positive energy plane
wave solutions of the Dirac (or Klein-Gordon) equation cannot have a finite
support \cite{thaller,hegerfeldt,hegerfeldt2}. Therefore the charge density computed from
the field operator will also exhibit infinite tails. In most practical
scenarios, the infinite tails of the wave packet can be neglected -- they are
very small, in particular for wave-packets that are wide enough (relative to
the Compton wavelength). But in some instances (e.g. when addressing issues
related to time inetrval detection), it might be desirable to avoid dealing with infinite
tails and model the initial wave-packet as having compact support.

In this paper, our aim is to extend this computational QFT scheme in order to
take into account wave-packets with initial compact support. The existence of
perfectly localized states in relativistic quantum mechanics and QFT remains a
controversial issue \cite{knight,shirikov,pav,witten,haag,horwitz,bracken} and
we will not dwell here into this debate. Our starting point is the pragmatic
observation that a state with compact support must contain both particle and
anti-particle components.\ Hence if such a state were measured right after
preparation, one would not obtain a particle with certainty, but might find an
anti-particle with a small probability. However in typical situations only the
particle component propagates towards a given direction of interest, at which point one
is dealing with the dynamics of a particle wave-packet without tails
interacting with the background field.

We will argue that the usual charge density operators cannot account for the
dynamics of compact support wave-packets, as for such wave-packets the local
charge is not well defined. We will build instead a density operator
reminiscent of the way in which states with compact support are treated in the
first quantized theory. By considering the pair creation process that occurs
when exciting the fermionic or bosonic vacua with a supercritical potential, the
density operator will allow us to study the propagation of these finite
support wave packets through arbitrary potentials.

The paper is organized as follows. In Sec. \ref{sec-prop}, we briefly recall
the propagation of wave packets with compact support in first quantized
relativistic quantum mechanics, and give a brief overview of wave-packet treatment
within the computational QFT framework we are using. We will see in
particular why the charge density operators fail to account for the
propagation of wave packets with finite support. In Sec. \ref{sec-blind} we
introduce a different density operator that considers modes associated with
both particles and anti-particles.\ This \textquotedblleft
charge-blind\textquotedblright\ density operator allows us to propagate wave
packets with compact support; we will examine some properties of the expectation values of this
operator. In
Sec. \ref{sec-ill} we illustrate both aspects of the density operator, first
by computing the free propagation of Dirac and Klein-Gordon wave packets
having compact support, and by carrying out numerical caclculations for such 
wave-packets scattering on a supercritical potential step giving rise to Klein tunneling. 
We close the paper
by discussing our findings and drawing our conclusions (Sec. \ref{sec-conc}).

\section{Propagation of wave packets}

\label{sec-prop} We recall here the propagation of wave packets in first
quantized relativistic quantum mechanics (RQM) and in the computational QFT
framework employed primarily to tackle the space-time resolved dynamics of bosonic or fermionic
fields in a background potential. We will more specifically focus on the
solutions of the Dirac and Klein-Gordon equations in one spatial dimension
with respective Hamiltonians $H^{D}=H_{0}^{D}+V$ and $H^{KG}=H_{0}^{KG}+V.$ In
the Dirac case, $H_{0}^{D}=-i\hbar c\alpha_{x}\partial_{x}+\beta mc^{2}$
($\alpha$ and $\beta$ are the usual Dirac matrices \footnote{We will consider
the usual one effective spatial dimension approximation, neglecting spin-flip
\cite{nitta} and replacing $\alpha_{x}$ and $\beta$ by the Pauli matrices
$\sigma_{1}$ and $\sigma_{3}$ respectively.}, $m$ the electron mass and $c$
the light velocity) while the free KG\ Hamiltonian is given by $H_{0}%
^{KG}=-\frac{\hbar^{2}}{2m}\left(  \tau_{3}+i\tau_{2}\right)  \partial_{x}%
^{2}+mc^{2}\tau_{3}$ (we are using the so-called Hamiltonian form of the KG
equation, where $\tau_{i}$ are the Pauli matrices and $m$ now represents the
boson mass \cite{greiner-rqm}). $V(x)$ is the background potential. The
eigenstates of the free Hamiltonians $H_{0}$ will be denoted by $|\phi
_{p}\rangle$ for positive energies and $|\varphi_{p}\rangle$ for negative
energies, where $\pm\left\vert E_{p}\right\vert =\pm\sqrt{p^{2}c^{2}%
+m^{2}c^{4}}$. We will use the same notation (including for the scalar
product) for the Dirac and Klein-Gordon cases and only specify differences
between bosons and fermions when relevant.

\subsection{Wave packets in the first quantized formalism}

In the first-quantized formalism a wave-packet with an arbitrary profile in
configuration space contains an expansion over both positive and negative
energy components \cite{greiner-rqm}. This is in particular the case for a
state having compact support. Let us assume $\psi(0,x)$ describes a state with
positive charge equal to 1 that is zero outside an interval $D$,%
\begin{equation}
\langle x|\psi\rangle=\psi(0,x)=N%
\begin{pmatrix}
G(x)\\
0
\end{pmatrix}
,
\end{equation}
where $G(x)$ is a function with compact support equal to zero outside $D$ and
$N$ is a normalization constant. We can rewrite this initial wave packet in
terms of the projection over the positive and negative energy components%
\begin{equation}
\psi(0,x)=\psi_{+}(0,x)+\psi_{-}(0,x)
\end{equation}
where we have defined%
\begin{equation}
\psi_{+}(0,x)=\int g_{+}(p)\langle x|\phi_{p}\rangle dp,\;\psi_{-}(0,x)=\int
g_{-}(p)\langle x|\varphi_{p}\rangle dp \label{1stWP}%
\end{equation}
and%
\begin{equation}
g_{+}(p)=\int dx\langle\phi_{p}|x\rangle\langle x|\psi\rangle,\ g_{-}(p)= -\epsilon  \int
dx\langle\varphi_{p}|x\rangle\langle x|\psi\rangle\
\end{equation}
where $\epsilon=1$ in the Klein-Gordon case and  $\epsilon=-1$ in the Dirac case.

At time $t$, this wave packet will have evolved to:%
\begin{equation}%
\begin{split}
\psi(t,x)  &  =\int g_{+}(p)\langle x|\phi_{p}\rangle e^{-iE_p t}dp+\int g_{-}(p)\langle x|\varphi_{p}\rangle e^{+iE_p t}dp\\
&  =\psi_{+}(t,x)+\psi_{-}(t,x)
\end{split}
\end{equation}
If the mean momentum of the initial wave packet is positive, the positive
energy part will propagate in the positive direction (towards the right) while
the negative energy one will propagate to the left (see eg \cite{our_dirac}
for an illustration). The density, which satisfies the continuity equation, is
given by%
\begin{equation}
r(t,x)=\psi^{\dagger}(t,x) \sigma\psi(t,x)=\psi_{+}(t,x)^{\dagger}%
\sigma\psi_{+}(t,x)+\psi_{-}(t,x)^{\dagger}\sigma\psi_{-}(t,x)+2\Re
\big(\psi_{+}^{\dagger}(t,x) \sigma\psi_{-}(t,x)\big) \label{frstq_dens}%
\end{equation}
where $\sigma$ is equal to the Pauli matrix $\tau_{3}$ in the case of bosons
and to the identity in the case of fermions. It is important to notice this
density can be interpreted as a probability density in the case of fermions
but for bosons it is a charge density that can be positive (particle
component) or negative (anti-particle component). Recall finally that the full
propagator (including positive as well as negative energy components) is
causal \cite{thaller}, so that a wave-packet with initial compact support will
always remain within the light-cone emanating from the bounds of the initial
support of the wavefunction.

\subsection{Computational QFT formalism}

\label{sec-cqft}

\subsubsection{Basic expressions}

We start from the familiar QFT expressions \cite{schweber} describing the
creation and annihilation of particles (`pa') and anti-particles (`an') at some
position $x$,
\begin{equation}%
\begin{split}
\hat{\Psi}_{pa}  &  =\int dp\hat{b}_{p}\langle x|\phi_{p}\rangle\\
\hat{\Psi}_{an}  &  =\int dp\hat{d}_{p}\langle x|\varphi_{p}\rangle^{\ast}%
\end{split}
\label{tife}%
\end{equation}
where $b_{p}$ and $d_{p}$ are the annihilation operators for a particle and an
anti-particle respectively (they yield a vanishing result when applied to the
vacuum). $b_{p}^{\dagger}$ and $d_{p}^{\dagger}$ are the corresponding
creation operators.\ These operators obey the usual commutation relations,
i.e. for fermions the only non-zero equal time anti-commutators are
$[b_{p},b_{k}^{\dagger}]_{+}=[d_{p},d_{k}^{\dagger}]_{+}=\delta(p-k)$ (and
similar commutators for bosons). We define as usual the conjugate of these
field operators acting on the dual Fock space
\begin{align}
\hat{\Psi}_{pa}^{\dagger}  &  =\int dp\hat{b}_{p}^{\dagger}\langle\phi
_{p}|x\rangle\label{tife2a}\\
\hat{\Psi}_{an}^{\dagger}  &  =\int dp\hat{d}_{p}^{\dagger}\langle\varphi
_{p}|x\rangle^{\ast} \label{tife2b}%
\end{align}
as well as their charge conjugate acting on the same Fock space,%
\begin{align}
\hat{\Psi}_{pa}^{\star}  &  =\int dp\hat{b}_{p}^{\dagger}\langle x|\phi
_{p}\rangle^{\ast}\label{tife3a}\\
\hat{\Psi}_{an}^{\star}  &  =\int dp\hat{d}_{p}^{\dagger}\langle x|\varphi
_{p}\rangle. \label{tife3b}%
\end{align}

The time evolution of these operators is obtained in the Heisenberg picture by
calculating the time evolution of each creation and annihilation operator
\cite{grobe-review} (see also the Supp.\ Mat. of \cite{ourqft}),%
\begin{equation}%
\begin{split}
b_{p}(t)  &  =\int dp^{\prime}\big(U_{\phi_{p}\phi_{p^{\prime}}}%
(t)b_{p^{\prime}}+U_{\phi_{p}\varphi_{p^{\prime}}}(t)d_{p^{\prime}}^{\dagger
}\big)\\
d_{p}^{\dagger}(t)  &  =\int dp^{\prime}\big(U_{\varphi_{p}\phi_{p^{\prime}}%
}(t)b_{p^{\prime}}+U_{\varphi_{p}\varphi_{p^{\prime}}}(t)d_{p^{\prime}%
}^{\dagger}\big)
\end{split}
\label{tdep_cre_ann}%
\end{equation}
where
\begin{equation}
U_{\zeta_{p}\xi_{q}}(t)=\langle\zeta_{p}|e^{-iHt}|\xi_{q}\rangle\label{ua}%
\end{equation}
are the evolution amplitudes generated by the full Hamiltonian (hence
including the background field). The field operators (\ref{tife}) therefore
become in the Heisenberg picture
\begin{equation}%
\begin{split}
\hat{\Psi}_{pa}(t,x)  &  =\int dpdp^{\prime}\big(U_{\phi_{p}\phi_{p^{\prime}}%
}(t)b_{p^{\prime}}+U_{\phi_{p}\varphi_{p^{\prime}}}(t)d_{p^{\prime}}^{\dagger
}\big)\langle x|\phi_{p}\rangle\\
\hat{\Psi}_{an}(t,x)  &  =\int dpdp^{\prime}\left(  U_{\varphi_{p}%
\phi_{p^{\prime}}}(t)b_{p^{\prime}}+U_{\varphi_{p}\varphi_{p^{\prime}}%
}(t)d_{p^{\prime}}^{\dagger}\right)  \langle x|\varphi_{n}\rangle
\end{split}
\label{tdfe}%
\end{equation}
with analog expressions for their conjugate.

The field operator obtained by quantizing the \textquotedblleft
classical\textquotedblright\ fields defined from the Klein-Gordon or Dirac
Lagrangians is related to the charge structure of the field \cite{haag}. It is
obtained by combining the operators given by Eqs (\ref{tife})-(\ref{tife3b}) as%
\begin{equation}
\hat{\Psi}(t,x)=\hat{\Psi}_{pa}(t,x)+\hat{\Psi}_{an}^{\star}(t,x)
\label{usualwo}%
\end{equation}
while the expression
\begin{equation}
\hat{\rho}_{ch}(t,x)=\hat{\Psi}^{\dagger}(t,x)\hat{\Psi}(t,x) \label{cdo}%
\end{equation}
represents the total charge density operator. Usually, one is interested in
the dynamics of the particle (or anti-particle) density.\ The particle density
operator is given by%
\begin{equation}
\hat{\rho}_{pa}(t,x)=\hat{\Psi}_{pa}^{\dagger}(t,x)\hat{\Psi}_{pa}(t,x).
\label{rhopa}%
\end{equation}
The particle density is then obtained as usual from the expectation value of
such operators.\ For instance denoting the vacuum by $\Vert0\rrangle$ (where
the symbol $\Vert\rrangle$ refers to a state in Fock space) the vacuum
expectation value $\llangle0\Vert\hat{\rho}_{pa}(t,x)\Vert0\rrangle$ gives the
space-time particle density (created by the background field) when there are
initially no particles nor anti-particles.\ In the present computational QFT
framework such densities are obtained by computing numerically the evolution
operator amplitudes given by Eq. (\ref{ua}) over a basis of solutions of the
free (Klein-Gordon or Dirac) Hamiltonian (see e.g. \cite{ourqft}). The total
number of positively charged particles is obtained as usual by integrating the
density,
\begin{equation}
N_{pa}(t)=\int dx\llangle0\Vert\hat{\rho}_{pa}(t,x)\Vert0\rrangle.
\end{equation}

The anti-particle density operator is defined from Eqs. (\ref{tife}%
)-(\ref{tife2b}) as
\begin{equation}
\hat{\rho}_{an}(t,x)=\hat{\Psi}_{an}^{\dagger}(t,x)\hat{\Psi}_{an}(t,x).
\label{rhoanti}%
\end{equation}
Note that the expectation values of this operator involves the scalar product
between the basis expansion functions $\varphi_{j}(t,x),$ which is positive
for Dirac fields but negative in the Klein-Gordon case.\ Hence the
anti-particle number is now given by%
\begin{equation}
N_{an}(t)=-\epsilon\int dx\llangle0\Vert\hat{\rho}_{an}(t,x)\Vert0\rrangle
\end{equation}
where $\epsilon=1$ for spin-0\ bosons and $\epsilon=-1$ for spin-1/2 fermions.

\subsubsection{Wave packet densities}

It is useful when considering a particle scattering on a potential, to model
the particle as a wave packet \ An initial particle wave packet $\chi(0,x)$ is
written in terms of creation operators $b_{p}^{\dagger}$ as
\cite{grobe-review}
\begin{equation}
\Vert\chi_+\rrangle=\int dpg_{+}(p)\langle x|\phi_{p}\rangle b_{p}^{\dagger
}\Vert0\rrangle. \label{iniqft}%
\end{equation}
The corresponding particle density is then given by the expectation value
\begin{equation}
\rho_{pa}(t,x)=\llangle\chi_+\Vert\hat{\rho}_{pa}(t,x)\Vert\chi_+\rrangle,
\label{rhopar}%
\end{equation}
representing the density created by the background field in the presence of
the wave-packet. The amplitudes $g_{+}(p)$ determine the spatial profile of the
wave packet, as is obvious by recalling that a first quantized
single-particle wavefunction $\chi(t,x)$ is obtained from the second quantized states as \cite{schweber}%
\begin{equation}
\chi(t,x)=\llangle0\Vert\hat{\Psi}(t,x)\Vert\chi\rrangle. \label{spw}%
\end{equation}
Note that if $\Vert\chi_+\rrangle$ of Eq. (\ref{iniqft}) is inserted into Eq. (\ref{spw}), the resulting wave-packet $\chi_+(t,x)$ contains only positive
energy modes, and can therefore only account for a particle wavefunction
 presenting infinite tails.\ This remains true if we replace Eq.
(\ref{iniqft}) by
\begin{equation}
\Vert\chi\rrangle=\int dp\big(g_{+}(p)b_{p}^{\dagger}+g_{-}(p)d_{p}^{\dagger
}\big)\Vert0\rrangle, \label{cs1}%
\end{equation}
which would be the analog of the first quantized wave-packet given by Eq. (\ref{1stWP}),
since the negative energy sector components vanish when inserted into Eq. (\ref{spw}). 

Similarly the single
particle wavefunction generated from the field operator $\hat{\Psi}^{\dagger
}(t,x)$ only keeps the negative energy modes, yielding an anti-particle
wavefunction with infinite tails that can only be approximately localized. We
therefore see that we cannot represent a wave-packet with compact support
within the computational QFT framework. Of course all the densities that can
be computed also present infinite tails -- $\rho_{pa}(t,x)$ projects to the
particle sector only and $\rho_{an}(t,x)=\llangle\chi\Vert\hat{\rho}%
_{an}(t,x)\Vert\chi\rrangle$ to the anti-particle sector, while the charge
density operator of Eq. (\ref{cdo}) can be seen to yield (see Appendix
\ref{appdx-a} for details) the charge density $\rho_{ch}(t,x)=\rho
_{pa}(t,x)-\epsilon\rho_{an}(t,x)$ (the tails in $\rho_{pa}$ and $\rho_{an}$
are different and do not cancel out).

\section{Density operator and localized wave-packets\label{sec-blind}}

\label{sec-nbden} In order to use the computational QFT formalism with wave
packets having compact spatial support, we need to define a density operator that does
not project to the particle or anti-particle sectors. This is done by going
back to Eqs. (\ref{tife}) and (\ref{tdfe}) and introducing%
\begin{equation}
\hat{\rho}(t,x)=\hat{\nu}^{\dagger}(t,x)\hat{\nu}(t,x)\label{rhodef}%
\end{equation}
where
\begin{equation}%
\begin{split}
\hat{\nu}^{\dagger}(t,x) &  =\hat{\Psi}_{pa}^{\dagger}(t,x)+\left(  \hat{\Psi
}_{an}^{\dagger}(t,x)\right)  ^{\ast}\\
\hat{\nu}(t,x) &  =\hat{\Psi}_{pa}(t,x)+\left(  \hat{\Psi}_{an}(t,x)\right)
^{\ast}.
\end{split}
\label{nudef}%
\end{equation}
The operator $\hat{\rho}(t,x)$ accounts for the density of particles and
anti-particles regardless of their charge. The rationale for taking this
combination of the field operators (\ref{tife}) is that rather than creating
(or annihilating) a charge, we are now creating (or destroying) a particle and
an anti-particle without changing the charge of a state. Indeed, the standard
field operator is well-known to be related to the charge structure in the
sense that $\Psi^{\dagger}$ raises the charge by 1, i.e. if $\hat{Q}=\int
dx\hat{\rho}_{ch}(x)$ is the total charge operator [see Eq. (\ref{cdo})] and
$\Vert q\rrangle$ is a state of charge $q$, i.e. $\hat{Q}\Vert
q\rrangle=q\Vert q\rrangle$, then $\hat{Q}\left[  \Psi^{\dagger
}\Vert q\rangle\rangle\right]  =(q+1)\left[  \Psi^{\dagger}\Vert
q\rrangle\right]  $ so that $\Psi^{\dagger}\Vert q\rrangle$
appears as a state of charge $q+1$. Similarly we can show that $\hat{\nu
}^{\dagger}$ increases the number of any $n$ particle $\Vert n\rrangle$
by 1: if $\hat{N}$ is the number operator, $\hat{N}=\hat{\rho}_{pa}%
-\epsilon\hat{\rho}_{an}=\int dp\left(  b_{p}^{\dagger}b_{p}^{{}}%
+d_{p}^{\dagger}d_{p}^{{}}\right)  $ and $\hat{N}\Vert n\rrangle=n\Vert
n\rrangle$, then we have (see Appendix \ref{Appdx-number})%
\begin{equation}
\hat{N}\left[  \nu^{\dagger}\Vert n\rangle\rangle\right]  =(n+1)\left[
\nu^{\dagger}\Vert n\rangle\rangle\right]  .\label{raising}%
\end{equation}
This implies in particular that $\hat{\rho}$ must contain terms allowing for
the conversion of a particle into an anti-particle (and vice versa). These are
the cross terms obtained when pluging-in Eq. (\ref{nudef}) into (\ref{rhodef}%
). 

As a consequence we can now accomodate a compact support wave-packet as given
by Eq. (\ref{cs1}) through
\begin{equation}
\chi(t,x)=\llangle0\Vert\hat{\nu}(t,x)\Vert\chi\rrangle\label{csr}%
\end{equation}
(compare with Eq. (\ref{spw})). In the presence of such a wave packet,
$\hat{\rho}(t,x)$ may be used to define a density that remains localized over
a compact support.\ Such a space-time resolved density is obtained from the
expectation value $\rho(t,x)=\llangle\chi\Vert\hat{\rho}_{n}(t,x)\Vert
\chi\rrangle$ which becomes%
\begin{equation}
\rho(x,t)=\llangle0\Vert\int dp\big(g_{+}^{\ast}(p)b_{p}+g_{-}^{\ast}%
(p)d_{p}\big)\hat{\rho}(t,x)\int dp\big(g_{+}(p)b_{p}^{\dagger}+g_{-}%
(p)d_{p}^{\dagger}\big)\Vert0\rangle\rangle.\label{rho}%
\end{equation}
In order to highlight the localized character of this density, we will parse
$\rho(t,x)$ as the density of particles and anti-particles created by the
background field on the one hand, and a wave packet density identical to the
first quantized single-particle wave-packet given by Eq. (\ref{frstq_dens}) on
the other, a wave-packet that is known to be supported on a compact
support. After some algebra (see Appendix \ref{appdx-b}) we obtain
\begin{equation}
\rho(t,x)=\rho_{1}(t,x)+\rho_{2}(t,x)+\rho_{3}(t,x).\label{rhotot}%
\end{equation}
The term%
\begin{equation}%
\begin{split}
\rho_{1}(t,x) &  =\int dp\big(\int dqU_{\phi_{p}\varphi_{q}}\langle x|\phi
_{p}\rangle\big)^{\dagger}\sigma\big(\int dqU_{\phi_{p}\varphi_{q}}\langle
x|\phi_{p}\rangle\big)\\
&  +\big(\int dpdqg_{+}(q)U_{\phi_{p}\phi_{q}}\langle x|\phi_{p}%
\rangle\big)^{\dagger}\sigma\big(\int dpdqg_{+}(q)U_{\phi_{p}\phi_{q}}\langle
x|\phi_{p}\rangle\big)\\
&  +\epsilon\Big(\int dpdqg_{-}^{\ast}(q)U_{\phi_{p}\varphi_{q}}\langle
x|\phi_{p}\rangle\Big)^{\dagger}\sigma\Big(\int dpdqg_{-}^{\ast}(q)U_{\phi
_{p}\varphi_{q}}\langle x|\phi_{p}\rangle\Big)
\end{split}
\label{rho1}%
\end{equation}
represents the density due to the presence of the background potential (first
line), the density corresponding to the incoming particle (second line), and
the modulation in the number density of the created particles due to the
incident particle wave packet. The structure od $\rho_{1}$ is identical to the particle density
defined by taking the expectation value of Eq. (\ref{rhopa}).

The second term, given by%
\begin{equation}%
\begin{split}
\rho_{2}(t,x)  &  =\int dp\left(  \int dqU_{\varphi_{p}\phi_{q}}\langle
x|\varphi_{p}\rangle^{\ast}\right)  ^{\dagger}\sigma\left(  \int
dqU_{\varphi_{p}\phi_{q}}\langle x|\varphi_{p}\rangle^{\ast}\right) \\
&  +\left(  \int dpdqg_{-}(q)U_{\varphi_{p}\varphi_{q}}\langle x|\phi
_{p}\rangle^{\ast}\right)  ^{\dagger}\sigma\left(  \int dpdqg_{-}%
(q)U_{\varphi_{p}\varphi_{q}}\langle x|\varphi_{p}\rangle^{\ast}\right) \\
&  +\epsilon\left(  \int dpdqg_{+}^{\ast}(q)U_{\varphi_{q}\varphi_{p}}\langle
x|\varphi_{p}\rangle^{\ast}\right)  ^{\dagger}\sigma\left(  \int
dpdqg_{+}^{\ast}(q)U_{\phi_{q}\varphi_{p}}\langle x|\varphi_{p}\rangle^{\ast
}\right)
\end{split}
\label{rho2}%
\end{equation}
is the counterpart of $\rho_{1}(t,x)$ for the anti-particle density and is
hence identical to the density obtained from the operator given by Eq.
(\ref{rhoanti}).

\begin{figure}[h]
	\includegraphics[scale=0.25]{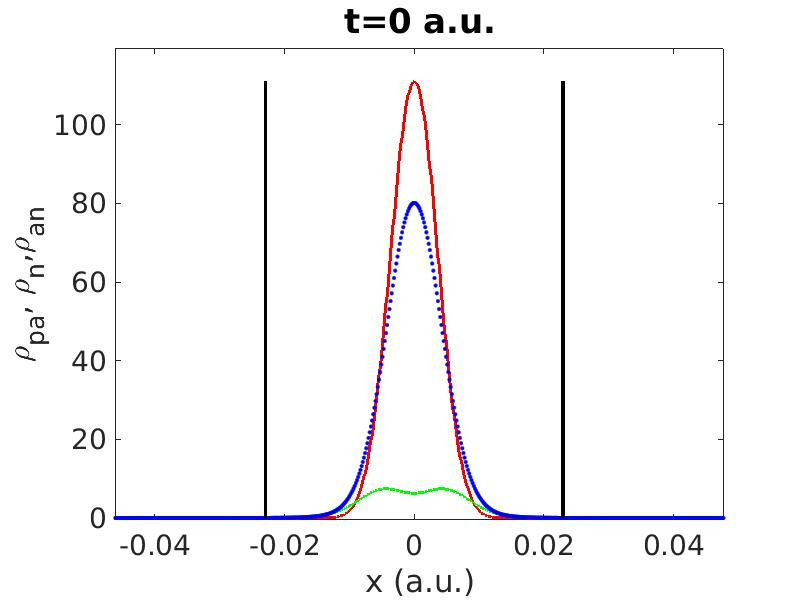}
	\includegraphics[scale=0.25]{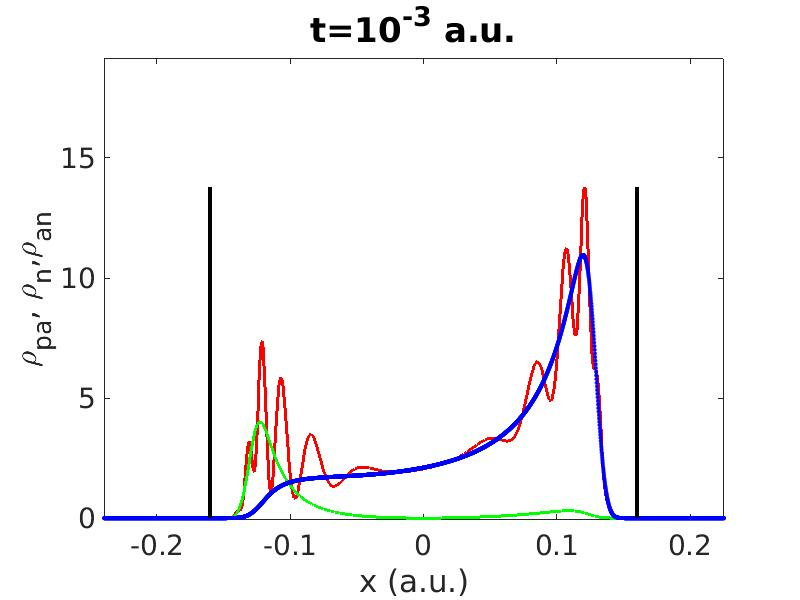}
	\includegraphics[scale=0.25]{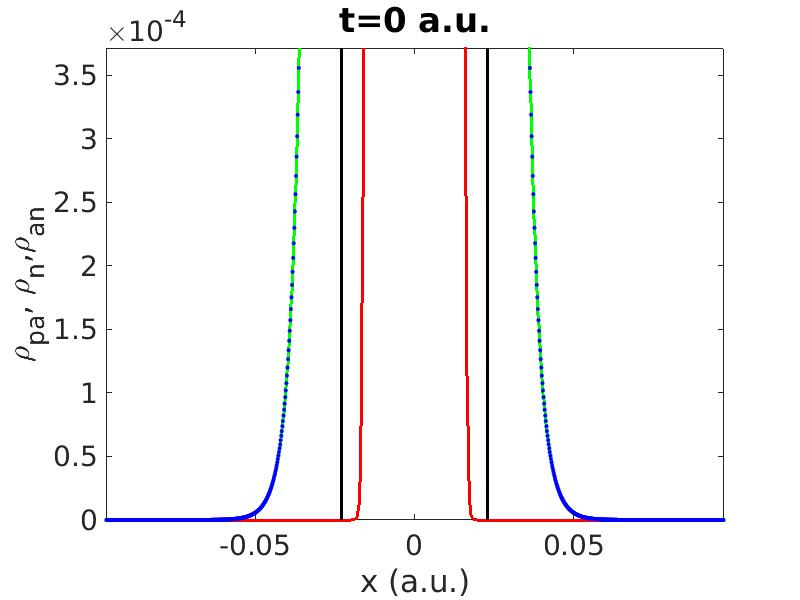}
	\includegraphics[scale=0.25]{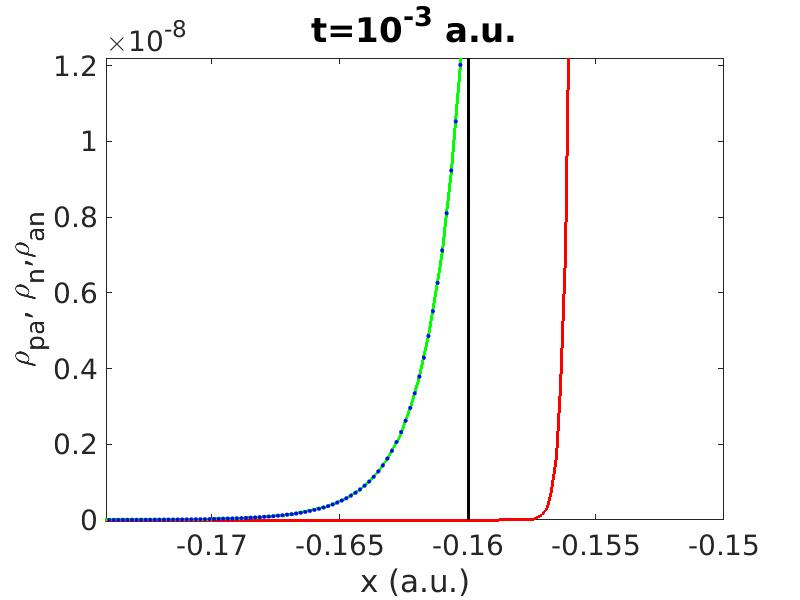}\caption{The spatial
		number densities for spin-1/2 fermions of unit mass (a. u.) are shown at $t=0$
		(left panels) and $t=10^{-3}$ a.u. (right panels). The bottom row zooms the
		top row densities to contrast the tails of the particle and anti-particle
		densities $\rho_{pa}$ (blue dotted) and $\rho_{an}$ (green) from the localized
		character of the proposed density $\rho$ (red line). In the left panel, the 2
		black lines indicate the bounds of the compact support density. In the right
		panel, the black lines are the positions of the light-cone emanating from
		these bounds. Note that a small fraction of the particle and anti-particle
		densities present tails outside the light-cone. The freely propagating wave packet is given by Eq. (\ref{eqwp}) with
		$D=2/c$ and $p_{0}=100$ a.u. (atomic units are used throughout).}%
	\label{fig_free_D}%
\end{figure}

Finally the third term in Eq. (\ref{rhotot}),%
\begin{equation}%
\begin{split}
\rho_{3}(t,x) &  =2\Re\Big(\int dpdqg_{-}^{\ast}(q)U_{\varphi_{p}\varphi_{q}%
}^{\ast}g_{+}(q)U_{\phi_{p}\phi_{q}}\langle\varphi_{q}|x\rangle\sigma\langle
x|\phi_{p}\rangle\Big)\\
&  +2\Re\Big(\int dpdqg_{-}^{\ast}(q)U_{\varphi_{p}\phi_{q}}^{\ast}%
g_{+}(q^{\prime})U_{\phi_{p}\varphi_{q}}\langle\varphi_{q}|x\rangle
\sigma\langle x|\phi_{p}\rangle\Big)
\end{split}
\label{rho3}%
\end{equation}
involves cross terms between positive and negative energy modes of the initial
wave-packet. This term accounts for the cancellation of the infinite spatial
tails intrinsic to $\rho_{1}$ and $\rho_{2}$. When integrated over the whole
space this term vanishes, ensuring that $\rho$ obeys%
\begin{equation}
\int dx\rho(t,x)=\int dx\rho_{pa}(t,x)+\int dx\rho_{an}(t,x).
\end{equation}
This is to be compared to the total number of particles $N$ given by the
expectation value of the number operator [cf.\ Eq. (\ref{numbo})] as%
\begin{equation}
N(t)=\int dx\rho_{pa}(t,x)-\epsilon\int dx\rho_{an}(t,x).\label{ntoteq}%
\end{equation}
We therefore see that for fermions ($\epsilon=-1)$ $\rho(t,x)$ can be
interpreted as a number density, whereas for bosons $\rho(t,x)$ appears as a
charge density.

Note that in the absence of a background field $\rho(t,x)$ represents the
evolution of the sole wave-packet.\ In this case, the density simplifies to
[put $U_{\phi_{p}\varphi_{q}}=0$ in Eq. (\ref{rhotot})]%
\begin{equation}%
\begin{split}
\rho^{(0)}(t,x) &  =\left(  \int\int dpdqg_{+}(q)U_{\phi_{p}\phi_{q}}\langle
x|\phi_{p}\rangle\right)  ^{\dagger}\sigma\left(  \int\int dpdqg_{+}%
(q)U_{\phi_{p}\phi_{q}}\langle x|\phi_{p}\rangle\right)  \\
&  +\left(  \int\int dpdqg_{-}(q)U_{\varphi_{p}\varphi_{q}}\langle x|\phi
_{p}\rangle\right)  ^{\dagger}\sigma\left(  \int\int dpdqg_{-}(q)U_{\varphi
_{p}\varphi_{q}}\langle x|\phi_{p}\rangle\right)  \\
&  +2\Re\Big(\int\int dpdqg_{-}(q)U_{\varphi_{p}\varphi_{q}}^{\ast}%
g_{+}(q^{\prime})U_{\phi_{p}\phi_{q}}\langle\varphi_{p}|x\rangle\sigma\langle
x|\phi_{p}^{\prime}\rangle\Big)
\end{split}
\label{equ0}%
\end{equation}
which is equal to the density calculated in the first quantized theory, given
by Eq. (\ref{frstq_dens}). This demonstrates that $\rho^{(0)}$ evolves within
a compact support inside the light cone. Moreover, in the absence of a
wave-packet, the density of created matter is equal to the sum of the two
number densities of particles and anti-particles since the cross terms
$\rho_{3}$ and all the terms with $g_{\pm}$ vanish in Eqs. (\ref{rho1}) and
(\ref{rho2}). This will be illustrated below.

\begin{figure}[h]
	\includegraphics[scale=0.25]{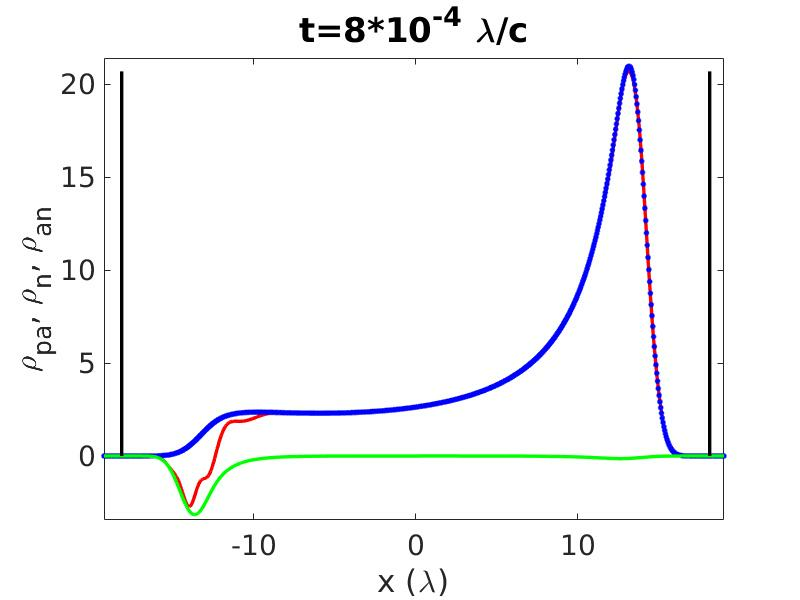}
	\includegraphics[scale=0.25]{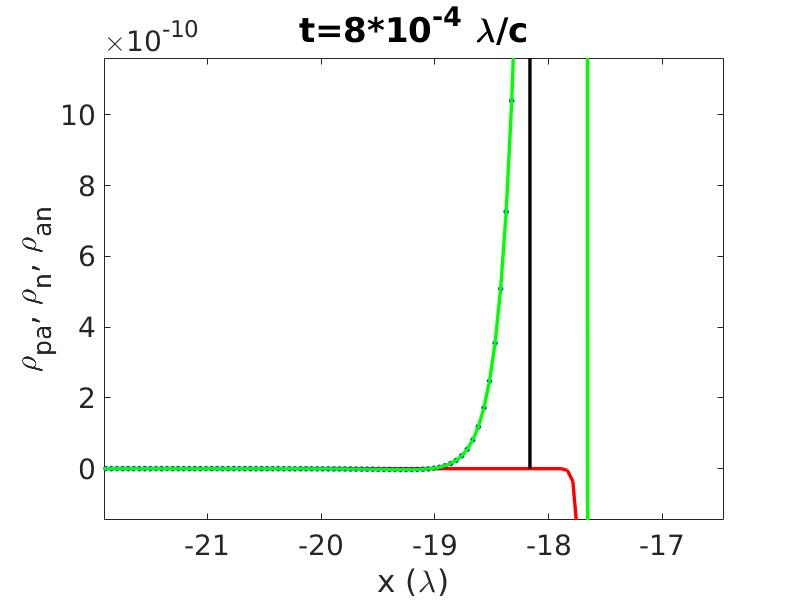}\caption{ Same as Fig.
		\ref{fig_free_D} but for a Klein-Gordon wave packet. Only the propagated
		wave-packet at $t=8\times10^{-4}$ is shown. The plot on the right panel is a
		zoom of the left panel figure near the position of the light-cone originating
		from the left bound of the $t=0$ compact support density. The color coding
		for the densities is the same as in Fig. \ref{fig_free_D}. Units are given in
		terms of the Compton wavelength $\lambda$. The freely propagating wave packet
		is given by Eq. (\ref{eqwp}) with $D=2\lambda$ and $p_{0}=100\hbar/\lambda$. }%
	\label{free_KG}%
\end{figure}

\section{Illustrations\label{sec-ill}}

\subsection{Free propagation of compact support wave packets}

We study numerically the time evolution of perfectly localized wave-packets of
a spin-0 boson or a spin-1/2 fermion using the bosonic or fermionic field
operators introduced above. The computational techniques employed, based on
accurate numerical computations of the densities on a finite space-time grid,
were detailed elsewhere (see Supp.\ Mat of Ref. \cite{ourqft}) for the case of the usual computational 
QFT framework; here we simply need to arrange the terms differently when computing $\rho(t,x)$.

\begin{figure}[h]

	\includegraphics[scale=0.3]{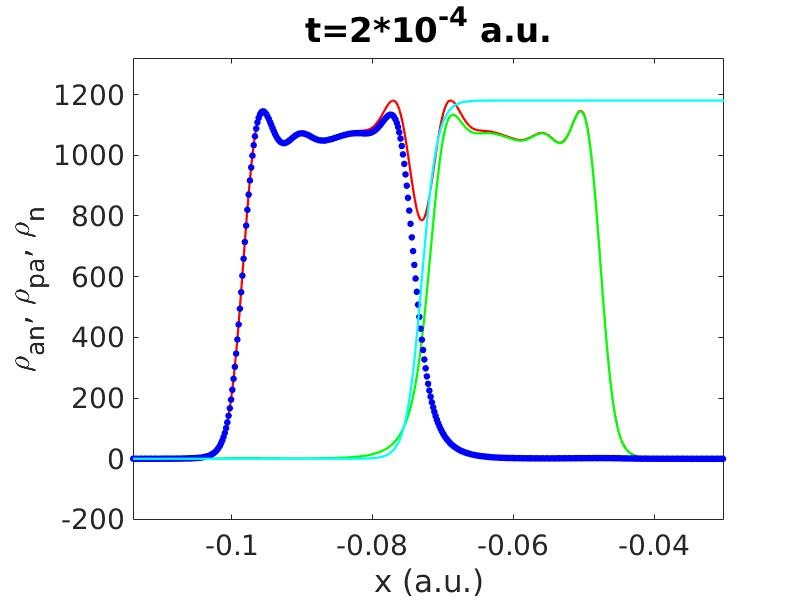}\caption{Vacuum expectation number for fermionic  particles (electrons) and anti-particles (positrons) created
		by a supercritical background field. The position of the step is represented in light blue, while $\rho_{pa}(t,x)$ and $\rho_{an}(t,x)$ 
		are pictured in dark blue and green resp. The vacuum expectation value of   $\hat{\rho}(t,x)$ (red line) appears as the algebraic sum $\rho_{pa}+\rho_{an}$ since $\rho_{3}$ [Eq. (\ref{rho3})] vanishes. The parameters of the step are $V_{0} = 9 m c^{2}$ and $d = - 10 \lambda$, with  $\alpha= 0.3/\lambda$. 	\label{fig_creat}}
\end{figure}

Let us take the initial wave packet%
\begin{equation}
\psi(x,0)=\mathcal{N}%
\begin{pmatrix}
G(x)\\
0
\end{pmatrix}
\label{eqwp}%
\end{equation}
where $\mathcal{N}$ is a normalization constant, and $G(x)$ is defined on the compact
support $x\in\lbrack x_{0}-D\pi/2,x_{0}+D\pi/2]$ as
\begin{equation}
G(x)=\cos^{8}(\frac{x-x_{0}}{D})e^{ip_{0}x}%
\end{equation}
This wave packet has a mean momentum $p_{0}$ and is centered around $x_{0}$ in
real space with a width $D$. We then determine the particle and anti-particle
densities employing the usual computational QFT framework introduced in Sec.
\ref{sec-cqft}, as well as the density operator proposed in Sec.
\ref{sec-blind}; for free propagation of interest in this sub-section, the
corresponding expression is given by Eq. (\ref{equ0}).

We show in Fig. \ref{fig_free_D} the fermionic (Dirac field) particle and
anti-particle densities $\rho_{pa}$ and $\rho_{an}$ as well as the density
$\rho$ for a freely propagating wave packet whose initial state is given by
Eq. (\ref{eqwp}) at two instants $t=0$ a.u. and $t=8\times10^{-4}$ a.u. The
particle as well as the anti-particle densities have infinite tails, while the
density $\rho$ reproduces the compact support of the initial wavefunction and
remains inside the light cone at later times.

Fig. \ref{free_KG} shows similar calculations for a bosonic (charged
Klein-Gordon) field. Only the evolved densities are shown (the initial
wave-packet, centered at $x_{0}=0$, is qualitatively similar to the one shown
in Fig. \ref{fig_free_D}). It can be seen that the particle density (blue
line) has moved towards the right while the anti-particle density (green line)
has moved towards the left. The density (red line) takes into account both
charges, and remains localized within the light-cone emanating from the
initial density, while the charged densities have infinite tails leaking from
the light cone.


\subsection{Klein tunneling with localized wave-packets}

Let us now examine the propagation of the densities in the presence of a
background field. For definiteness, let us take a supercritical potential step
rising at $x=d$ 
given by:
\begin{equation}
V(x)= V_{0}(1+\tanh\left(  (x-d)/\alpha\right))/2  ,\label{pot}%
\end{equation}
where $V_0$ is the step height and $\alpha$ the smoothness parameter of the background field. In the absence of any wave-packet, the background potential creates
particle/anti-particle pairs. This is illustrated in Fig. \ref{fig_creat}
where the space-time resolved densities $\rho_{pa},$  $\rho_{an}$ and $\rho$
are plotted in the fermionic case.

\begin{figure}[h]
 \includegraphics[scale=0.25]{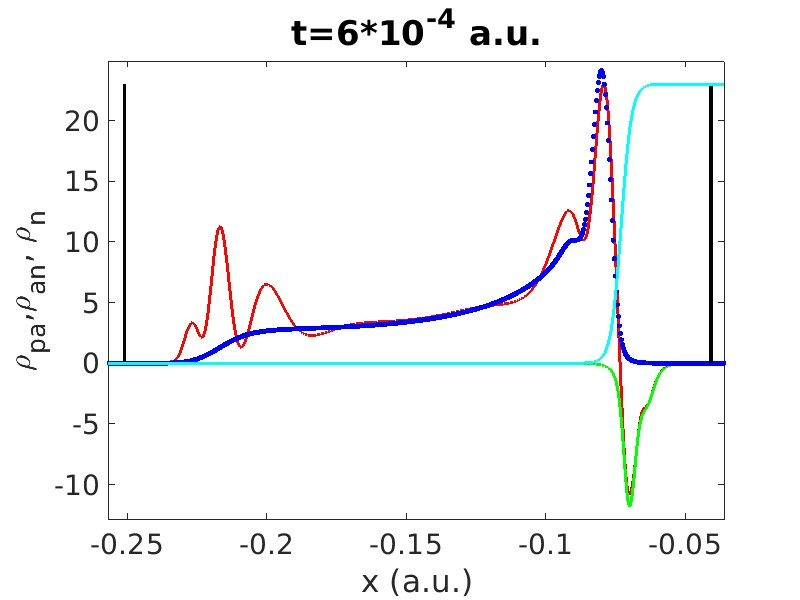}
\includegraphics[scale=0.25]{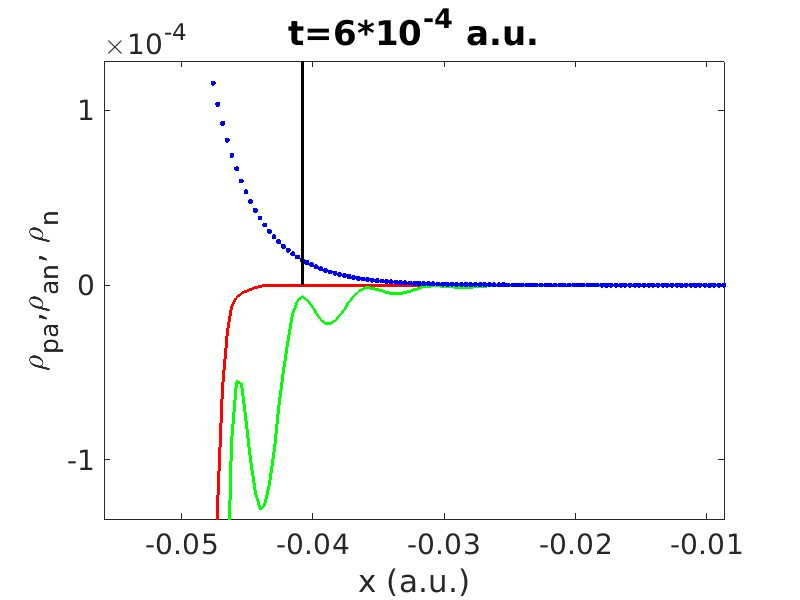}
\caption{Wave-packet densities (obtained by subtracting the terms corresponding to pair creation) for an electron packet scattering on a supercritical potential step of height
	$V_{0} = 9 mc^{2}$, smoothness parameter $\alpha= 0.3/c$, and centered around
	$d=-10 \lambda$. The particle (electron) density is shown in blue, the anti-particle (positron) density in green
and the compact support density in red.
The plot on the right zooms on the region around the light-cone originating at the $t=0$ right boundary
of the spatial support of the initial wave-packet: the
compact-support density is seen to vanish outside the light-cone while this is not the case for the particle and anti-particle densities.}%
\label{fig_step_D}
\end{figure}

The more interesting case is that of Klein tunneling, in which a wave-packet
scatters on the supercritical step and propagates undamped in the potential
region. The field operators now account for pair creation and wave-packet
propagation.\ In the standard computational QFT treatment
\cite{grobe-review,ourqft}, the wave-packet displays infinite tails at any
time, whereas the densities proposed in Sec. \ref{sec-blind} remain within the
light-cone emanating from the compact support region over which the initial
wave-packet is localized. This is illustrated in Fig. \ref{fig_step_D}. We plot there, for the fermionic
case, only the terms in $\rho_{1}$, $\rho_{2}$ and $\rho_3$ [cf. Eqs. (\ref{rhotot})-(\ref{rho3})] containing
terms relevant to the wave-packet (which is tantamount to substracting the terms in the total density 
that account for pair creation).

\section{Discussion and conclusion}

\label{sec-conc}

We have proposed in this paper a way to work with states defined over a
compact spatial support in the framework of computational QFT. Our approach hinged on
introducing a density operator $\hat{\rho}(t,x)$ [Eq. (\ref{rhodef})] from
which expectation values yield densities lacking the infinite tails
characterizing the standard computational QFT formalism. Our starting point was the remark that
a compact support KG or Dirac wave packet must have both positive and negative
energy components: by relying on the connection between QFT states and first
quantized ones, we have seen how such a density operator could be constructed.
We have analyzed some properties of the expectation values of the proposed
operator, and computed numerical results for specific illustrations for
Klein-Gordon and Dirac fields (free propagation of a wave-packet, and a
wave-packet impinging on a Klein step).

We emphasize that the approach developed here should be regarded as a
practical recipe in order to manipulate compactly localized states 
when studying space-time resolved wave-packet dynamical problems in
situations in which it is awkward  to use the ``essentially
localized states'' \cite{haag}  of standard quantum field
theory intrinsically displaying infinite tails. Such states 
could be useful for example in certain detector models aiming at measuring arrival
times. We are not claiming that the operators $\nu(x,t)$ and $\nu^{\dagger
}(x,t)$ of Eq. (\ref{nudef}) can be promoted to fundamental quantities from
which a full-fledged quantum field approach can be defined. Previous attempts to construct 
strictly localized states from vacuum excitations of a quantized field have run into difficulties, 
such as field-theoretic Hamiltonians unbounded from below \cite{shirikov}. 
Nevertheless, besides the practical usage alluded to above, the present results could also be useful in investigating problems such as optimizing
the localization properties of a single particle state, or conversely finding how close a strictly localized
wave-packet can be to a genuine single particle state. These problems will be investigated in the future.

\bigskip

\appendix

\section{Densities with infinite tails in the standard computational QFT
framework\label{appdx-a}}

Let us determine the particle density $\rho_{pa}(t,x)$ in the absence of a
background potential. $\rho_{pa}$, defined by Eq. (\ref{rhopar}), becomes%
\begin{equation}
\rho_{pa}(t,x)=\llangle0\Vert\int dp\big(g_{+}^{\ast}(p)b_{p}+g_{-}^{\ast
}(p)d_{p}\big)\hat{\rho}_{pa}(t,x)\int dp\big(g_{+}(p)b_{p}^{\dagger}%
+g_{-}(p)d_{p}^{\dagger}\big)\Vert0\rrangle \label{par_den_ope}%
\end{equation}
Writing $\hat{\rho}_{pa}(t,x)$ in terms of the field operators, Eq.
(\ref{rhopa}), using Eq.(\ref{tdep_cre_ann}), and noticing that $U_{\phi
_{p}\varphi_{p^{\prime}}}=0$ in the case of free propagation, one obtains
\begin{equation}%
\begin{split}
\rho_{pa}(t,x)=\llangle0\Vert\int dp\big(g_{+}^{\ast}(p)b_{p}+g_{-}^{\ast
}(p)d_{p}\big)  &  \iint dpdp^{\prime}U_{\phi_{p}\phi_{p^{\prime}}}^{\ast
}b_{p^{\prime}}^{\dagger}\langle\phi_{p}|x\rangle\\
&  \iint dpdp^{\prime}U_{\phi_{p}\phi_{p^{\prime}}}b_{p^{\prime}}\langle
x|\phi_{p}\rangle\\
\  &  \int dp\big(g_{+}(p)b_{p}^{\dagger}+g_{-}(p)d_{p}^{\dagger}%
\big)\Vert0\rrangle
\end{split}
\end{equation}
The only non-vanishing term in this case gives:%
\begin{equation}
\rho_{pa}(t,x)=\left(  \int dpg_{+}(p)U_{\phi_{p^{\prime}}\phi_{p}}\langle
x|\phi_{p}\rangle\right)  ^{\dagger}\sigma\left(  \int dpg_{+}(p)U_{\phi
_{p^{\prime}}\phi_{p}}\langle x|\phi_{p}\rangle\right)
\end{equation}
Similarly, the density of anti-particles is computed as
\begin{equation}
\rho_{an}(t,x)=\left(  \int dpg_{+}(p)U_{\varphi_{p^{\prime}}\varphi_{p}%
}\langle x|\varphi_{p}\rangle^{\ast}\right)  ^{\dagger}\sigma\left(  \int
dpg_{+}(p)U_{\varphi_{p^{\prime}}\varphi_{p}}\langle x|\varphi_{p}%
\rangle^{\ast}\right)  \label{rhoan}%
\end{equation}

Now using,
\begin{equation}%
\begin{split}
U_{\phi_{p}\phi_{q}}  &  =\int dx\langle\phi_{p}|x\rangle\sigma\langle
x|e^{-iE_{q}t}|\phi_{q}\rangle=e^{-iE_{q}t}\int dx\langle\phi_{p}%
|x\rangle\sigma\langle x|\phi_{q}\rangle\\
U_{\varphi_{p}\varphi_{q}}  &  =\int dx\langle\varphi_{p}|x\rangle
\sigma\langle x|e^{-iE_{q}t}|\varphi_{q}\rangle=e^{iE_{q}t}\int dx\langle
\varphi_{p}|x\rangle\sigma\langle x|\varphi_{q}\rangle
\end{split}
\label{freeU}%
\end{equation}
where
\begin{equation}%
\begin{split}
\int dx\langle\phi_{p}|x\rangle\sigma\langle x|\phi_{q}\rangle &  =\delta
_{pq}\\
\int dx\langle\varphi_{p}|x\rangle\sigma\langle x|\varphi_{q}\rangle &
=-\epsilon\delta_{pq}%
\end{split}
\end{equation}
where $\epsilon=+1(-1)$ in the bosonic (fermionic) case, one obtains:
\begin{align}
\rho_{pa}(t,x)  &  =\left(  \int dpg_{+}(q)e^{-iE_{p}}\langle x|\phi
_{p}\rangle\right)  ^{\dagger}\sigma\left(  \int dpg_{+}(q)e^{-iE_{p}}\langle
x|\phi_{p}\rangle\right) \label{apa}\\
\rho_{an}(t,x)  &  =\left(  \int dpg_{-}(q)e^{iE_{p}}\langle x|\varphi
_{p}\rangle\right)  ^{\dagger}\sigma\left(  \int dpg_{-}(q)e^{iE_{p}}\langle
x|\varphi_{p}\rangle\right)  \label{ana}%
\end{align}
It is straightforward to see that these densities are exactly equal to the
first quantized densities $\psi_{+}(t,x)^{\dagger}\sigma\psi_{+}(t,x)$ and
$\psi_{-}(t,x)^{\dagger}\sigma\psi_{-}(t,x)$ respectively (see Eq.
(\ref{1stWP})). They can thus not be localized within a region of compact
support, given that the cross term in Eq. (\ref{frstq_dens}) is necessary to
suppress the tails.

The total charge density $\rho_{ch}(t,x)$, defined by Eq. (\ref{cdo}) and
whose interation over all space defines the usual charge operator
\cite{peskin,greiner,schweber}, also presents tails. This can be seen
immediately by relying on the textbook result
\begin{equation}
\rho_{ch}(x,t)=\rho_{pa}(t,x) - \epsilon\rho_{an}(t,x)
\end{equation}
and using the reasoning below Eqs. (\ref{apa}-(\ref{ana}). Alternatively it
can be directly derived within the present framework, satrting from the field
operator (\ref{usualwo}) in the Heisenberg picture,
\begin{equation}
\hat{\Psi}(t,x)=\int dp\big(b_{p}^{\dagger}(t)\langle x|\phi_{p}\rangle
+d_{p}(t)\langle x|\varphi_{p}\rangle\big). \label{chfiop}%
\end{equation}
Eq. (\ref{cdo}) leads to%
\begin{equation}%
\begin{split}
\rho_{ch}(t,x)=\llangle0\Vert\int dp\big(g_{+}^{\ast}(p)b_{p}+g_{-}^{\ast
}(p)d_{p}\big)  &  \iint dpdp^{\prime}\Big(U_{\phi_{p}\phi_{p^{\prime}}}%
^{\ast}b_{p^{\prime}}^{\dagger}\langle\phi_{p}|x\rangle+U_{\varphi_{p}%
\varphi_{p^{\prime}}}^{\ast}d_{p^{\prime}}\langle\varphi_{p}|x\rangle\Big)\\
&  \iint dpdp^{\prime}\Big(U_{\phi_{p}\phi_{p^{\prime}}}b_{p^{\prime}}%
\langle\phi_{p}|x\rangle+U_{\varphi_{p}\varphi_{p^{\prime}}}d_{p^{\prime}%
}^{\dagger}\langle x|\varphi_{p}\rangle\Big)\\
&  \int dp\big(g_{+}(p)b_{p}^{\dagger}+g_{-}(p)d_{p}^{\dagger}\big)\Vert
0\rrangle
\end{split}
\end{equation}
which can be simplified to%
\begin{equation}%
\begin{split}
\rho_{ch}(x,t)  &  =-\epsilon\left(  \iint dpdqg_{-}(q)U_{\varphi_{p}%
\varphi_{q}}\langle x|\varphi_{p}\rangle\right)  ^{\dagger}\sigma\left(  \iint
dpdqg_{-}(q)U_{\varphi_{p}\varphi_{q}}\langle x|\varphi_{p}\rangle\right) \\
&  +\left(  \iint dpdqg_{+}(q)U_{\phi_{p}\phi_{q}}\langle x|\phi_{p}%
\rangle\right)  ^{\dagger}\sigma\left(  \iint dpdqg_{+}(q)U_{\phi_{p}\phi_{q}%
}\langle x|\phi_{p}\rangle\right)  .
\end{split}
\end{equation}

\section{Number raising operator\label{Appdx-number}}

We provide here a proof of Eq. (\ref{raising}). 

Let%
\begin{equation}
\hat{N}=\int dp\left(  \hat{b}_{p}^{\dagger}\hat{b}_{p}+d_{p}^{\dagger}%
d_{p}\right)  \label{numbo}%
\end{equation}
denote the total particle number. From Eq. (\ref{nudef}), $\hat{\nu}^{\dagger
}$ can be written as%
\begin{equation}
\hat{\nu}^{\dagger}=\int dp^{\prime}\left(  \hat{b}_{p^{\prime}}^{\dagger}%
\phi_{p^{\prime}}+\hat{d}_{p^{\prime}}^{\dagger}\varphi_{p^{\prime}}\right)  .
\end{equation}
Therefore, we have%
\begin{equation}
\hat{N}\hat{\nu}^{\dagger}=\int\int dpdp^{\prime}\left(  \hat{b}_{p}^{\dagger
}\hat{b}_{p}+\hat{d}_{p}^{\dagger}\hat{d}_{p}\right)  \left(  \hat
{b}_{p^{\prime}}^{\dagger}\phi_{p^{\prime}}+\hat{d}_{p^{\prime}}^{\dagger
}\varphi_{p^{\prime}}\right)
\end{equation}
and by using the fermionic or bosonic commutation relations in the first and
last terms we obtain%
\begin{equation}%
\begin{split}
\hat{N}\hat{\nu}^{\dagger} &  =\int\int dpdp^{\prime}\left(  \epsilon\hat
{b}_{p}^{\dagger}\hat{b}_{p^{\prime}}^{\dagger}\hat{b}_{p}\phi_{p^{\prime}%
}+\delta(p-p^{\prime})\hat{b}_{p}^{\dagger}\phi_{p^{\prime}}+\hat{d}%
_{p}^{\dagger}\hat{d}_{p}\hat{b}_{p^{\prime}}^{\dagger}\phi_{p^{\prime}%
}\right)  \\
&  +\int\int dpdp^{\prime}\left(  \hat{b}_{p}^{\dagger}\hat{b}_{p}\hat
{d}_{p^{\prime}}^{\dagger}\varphi_{p^{\prime}}+\epsilon\hat{d}_{p}^{\dagger
}\hat{d}_{p^{\prime}}^{\dagger}\hat{d}_{p}\varphi_{p^{\prime}}+\delta
(p-p^{\prime})\hat{d}_{p}^{\dagger}\varphi_{p^{\prime}}\right)  .
\end{split}
\end{equation}
Integrating the Dirac deltas and using again the the commutativity
(anti-commutativity) of bosons (fermions) in the first and forth terms leads
to%
\begin{equation}%
\begin{split}
\hat{N}\hat{\nu}^{\dagger} &  =\int dp^{\prime}\hat{b}_{p^{\prime}}^{\dagger
}\phi_{p^{\prime}}\int dp\left(  \hat{b}_{p}^{\dagger}\hat{b}_{p}+\hat{d}%
_{p}^{\dagger}\hat{d}_{p}\right)  +\int dp^{\prime}\hat{b}_{p^{\prime}%
}^{\dagger}\phi_{p^{\prime}}\\
&  +\int dp^{\prime}\hat{d}_{p^{\prime}}^{\dagger}\varphi_{p^{\prime}}\int
dp\left(  \hat{b}_{p}^{\dagger}\hat{b}_{p}+\hat{d}_{p}^{\dagger}\hat{d}%
_{p}\right)  +\int dp^{\prime}\hat{d}_{p^{\prime}}^{\dagger}\varphi
_{p^{\prime}}%
\end{split}
\end{equation}
and hence%
\begin{equation}%
\begin{split}
\hat{N}\hat{\nu}^{\dagger} &  =\int dp^{\prime}\hat{b}_{p^{\prime}}^{\dagger
}\phi_{p^{\prime}}(\hat{N}+1)+\int dp^{\prime}\hat{d}_{p^{\prime}}^{\dagger
}\varphi_{p^{\prime}}(\hat{N}+1)\\
&  =\hat{\nu}^{\dagger}(\hat{N}+1)
\end{split}
\end{equation}
from which Eq. (\ref{raising}) follows.

\section{Densities with compact support}

\label{appdx-b} Let us compute the expectation value of the density defined by
Eq. (\ref{rhodef}) in the presence of an initial wave-packet with compact
support.\ The resulting expression [Eq. (\ref{rho})] becomes, by inserting Eq.
(\ref{tdep_cre_ann}) :
\begin{equation}%
\begin{split}
\rho(t,x)=  &  \langle\langle0\Vert\int dp\big(g_{+}^{\ast}(p)b_{p}%
+g_{-}^{\ast}(p)d_{p}\big)\Big\{\int dp_{1}dp_{2}\phi_{p_{1}}^{\dagger}%
\phi_{p_{2}}b_{p_{1}}^{\dagger}(t)b_{p_{2}}(t)\\
+  &  \int dp_{1}dp_{2}\varphi_{p_{1}}^{\dagger}\varphi_{p_{2}}d_{p_{1}%
}^{\dagger}(t)d_{p_{2}}(t)+\Big(\int dp_{1}dp_{2}\phi_{p_{1}}^{\dagger}%
\varphi_{p_{2}}b_{p}^{\dagger}(t)d_{p}(t)+HC\Big)\Big\}\\
&  \int dp\big(g_{+}(p)b_{p}^{\dagger}+g_{-}(p)d_{p}^{\dagger}\big)\Vert
0\rrangle
\end{split}
\label{rho_ex}%
\end{equation}
We then parse these terms as per Eq. (\ref{rhotot}), where%
\begin{equation}%
\begin{split}
\rho_{1}(t,x)=\llangle0\Vert &  \int dp\big(g_{+}^{\ast}(p)b_{p}+g_{-}^{\ast
}(p)d_{p}\big)\Big\{\int dp_{1}dp_{2}\langle\phi_{p_{1}}|x\rangle\langle
x|\phi_{p_{2}}\rangle\int dp^{\prime}\Big(U^{*}_{\phi_{p_{1}}\phi_{p^{\prime}%
}}(t)b_{\phi_{p^{\prime}}}^{\dagger}+U^{*}_{\phi_{p_{1}}\varphi_{p^{\prime}}%
}(t)d_{\varphi_{p^{\prime}}}\Big)\\
&  \int dp^{\prime}\Big(U_{\phi_{p_{2}}\phi_{p^{\prime}}}(t)b_{\phi
_{p^{\prime}}}+U_{\phi_{p_{2}}\varphi_{p^{\prime}}}(t)d^{\dagger}%
_{\varphi_{p^{\prime}}}\Big)\Big\}\int dp\big(g_{+}(p)b_{p}^{\dagger}%
+g_{-}(p)d_{p}^{\dagger}\big)\Vert0\rrangle\\
&
\end{split}
\end{equation}
becomes by applying the creation and annhilation operators to the vacuum state%
\begin{equation}%
\begin{split}
\rho_{1}(t,x)  &  =\llangle0\Vert\int dq_{1}dq_{1}^{\prime}dq_{2}%
dq_{2}^{\prime}dp_{1}dp_{2}g_{-}^{\ast}(q_{1})g_{-}(q_{2})U_{\phi_{p_{1}%
}\varphi_{q_{1}^{\prime}}}^{\ast}(t)U_{\phi_{p_{2}}\varphi_{q_{2}^{\prime}}%
}(t)\langle\phi_{p_{1}}|x\rangle\langle x|\phi_{p_{2}}\rangle d_{q_{1}%
}d_{q_{1}^{\prime}}d_{q_{2}^{\prime}}^{\dagger}d_{q_{2}}^{\dagger}%
\Vert0\rrangle\\
&  +\llangle0\Vert\int dq_{1}dq_{1}^{\prime}dq_{2}dq_{2}^{\prime}dp_{1}%
dp_{2}g_{+}^{\ast}(q_{1})g_{+}(q_{2})U_{\phi_{p_{1}}\varphi_{q_{1}^{\prime}}%
}^{\ast}(t)U_{\phi_{p_{2}}\varphi_{q_{2}^{\prime}}}(t)\langle\phi_{p_{1}%
}|x\rangle\langle x|\phi_{p_{2}}\rangle b_{q_{1}}d_{q_{1}^{\prime}}%
d_{q_{2}^{\prime}}^{\dagger}b_{q_{2}}^{\dagger}\Vert0\rrangle\\
&  +\llangle0\Vert\int dq_{1}dq_{1}^{\prime}dq_{2}dq_{2}^{\prime}dp_{1}%
dp_{2}g_{+}^{\ast}(q_{1})g_{+}(q_{2})U_{\phi_{p_{1}}\phi_{q_{1}^{\prime}}%
}^{\ast}(t)U_{\phi_{p_{2}}\phi_{q_{2}^{\prime}}}(t)\langle\phi_{p_{1}%
}|x\rangle\langle x|\phi_{p_{2}}\rangle b_{q_{1}}b_{q_{1}^{\prime}}^{\dagger
}b_{q_{2}^{\prime}}b_{q_{2}}^{\dagger}\Vert0\rrangle.\\
&
\end{split}
\end{equation}
Using
\begin{equation}%
\begin{split}
&  \llangle0\Vert d_{q_{1}}d_{q_{1}^{\prime}}d_{q_{2}^{\prime}}^{\dagger
}d_{q_{2}}^{\dagger}\Vert0\rrangle=\delta_{q_{1}^{\prime}q_{2}^{\prime}}%
\delta_{q_{1}q_{2}}+\epsilon\delta_{q_{1}q_{2}^{\prime}}\delta_{q_{1}^{\prime
}q_{2}}\\
&  \llangle0\Vert b_{q_{1}}d_{q_{1}^{\prime}}d_{q_{2}^{\prime}}^{\dagger
}b_{q_{2}}^{\dagger}\Vert0\rrangle=\delta_{q_{1}q_{2}}\delta_{q_{1}^{\prime
}q_{2}^{\prime}}\\
&  \llangle0\Vert b_{q_{1}}b_{q_{1}^{\prime}}b_{q_{2}^{\prime}}^{\dagger
}b_{q_{2}}^{\dagger}\Vert0\rrangle=\delta_{q_{1}q_{2}^{\prime}}\delta
_{q_{2}q_{2}^{\prime}},
\end{split}
\end{equation}
one obtains:%
\begin{equation}%
\begin{split}
\rho_{1}(t,x)  &  =\epsilon\int dq\big|g_{-}(q)\big|^{2}\int dq\left(  \int
U_{\phi_{p}\varphi_{q}}(t)\langle x|\phi_{p}\rangle\right)  ^{\dagger}%
\sigma\left(  \int U_{\phi_{p}\varphi_{q}}(t)\langle x|\phi_{p}\rangle\right)
\\
&  +\int dq\big|g_{+}(q)\big|^{2}\int dq\left(  \int U_{\phi_{p}\varphi_{q}%
}(t)\langle x|\phi_{p}\rangle\right)  ^{\dagger}\sigma\left(  \int U_{\phi
_{p}\varphi_{q}}(t)\langle x|\phi_{p}\rangle\right) \\
&  +\left(  \int dpdpg_{+}(p)U_{\phi_{p}\phi_{q}}\langle x|\phi_{p}%
\rangle\right)  ^{\dagger}\sigma\left(  \int dpdpg_{+}(p)U_{\phi_{p}\phi_{q}%
}\langle x|\phi_{p}\rangle\right) \\
&  +\epsilon\left(  \int dpdpg_{-}^{\ast}(p)U_{\phi_{p}\varphi_{q}}\langle
x|\phi_{p}\rangle\right)  ^{\dagger}\sigma\left(  \int dpdpg_{-}^{\ast
}(p)U_{\phi_{p}\varphi_{q}}\langle x|\phi_{p}\rangle\right)
\end{split}
\end{equation}
Using the normalization of the initial wave packet,
\begin{equation}
\int dq\big|g_{+}(q)\big|^{2}-\epsilon\int dq\big|g_{-}(q)\big|^{2}=1
\end{equation}
yields Eq. (\ref{rho1}). The terms $\rho_{2}(t,x)$ and $\rho_{3}(t,x)$ are obtained\ similarly.


\begin{thebibliography}{99}                                                                                               %


\bibitem {schw}J. Schwinger. 
Phys. Rev. 82:664-679, (1951).

\bibitem {review}A. Fedotov, A. Ilderton, F. Karbstein, B. King, D. Seipt, H. Taya and G. Torgrimsson, Phys. Rep. 1010, 1-138 (2023).

\bibitem {gitman}S. P. Gavrilov and D. M. Gitman, Phys. Rev. D 93, 045002, (2016).

\bibitem {grobe-review}T. Cheng, Q. Su, and R. Grobe,
Contemp. Phys. 51, 315 (2010).

\bibitem {grobe_bo}R. E. Wagner, M. R. Ware, Q. Su, and R. Grobe Phys. Rev. A
81, 024101 (2010).

\bibitem {hansen}A. Hansen and F. Ravndal, Phys. Scr. 23, 1036 (1981).

\bibitem {nakazato}H. Nakazato and M. Ochiai, Prog. Theor. Exp. Phys.  073B02 (2022).



\bibitem {grobe_rates}T. Cheng, M. R. Ware, Q. Su, and R. Grobe Phys. Rev. A
80, 062105 (2009)

\bibitem {grobe_arb}T. Cheng, Q. Su, and R. Grobe, Phys. Rev. A 80, 013410 (2009).

\bibitem {wang}Q. Wang, J. Liu and L. B.  Fu Scientific Rep.  6 25292 (2016)

\bibitem {lv}Q. Z. Lv and H. Bauke, Phys. Rev. D 96, 056017 (2017)

\bibitem {ourqft}M. Alkhateeb and A. Matzkin, Phys. Rev. A 106, L060202, (2022).

\bibitem {thaller}B. Thaller. The Dirac Equation (Springer, Berlin, 2009),
Chap 1.

\bibitem {hegerfeldt}G. C. Hegerfeldt, Phys. Rev. D 10, 3320 (1974)

\bibitem {hegerfeldt2}G. C.
Hegerfeldt and S. N. M. Ruijsenaars, Phys. Rev. D 22, 377 (1980)

\bibitem {bracken}A. J. Bracken and G. F. Melloy, 
J.Phys. A 32, 6127 (1999).

\bibitem {witten}E. Witten, Rev. Mod. Phys. 90, 045003, (2018).

\bibitem {knight}J.\ M.\ Knight, J. Math. Phys. 2, 459 (1961).

\bibitem {shirikov}M. I. Shirokov, Theor. Math. Phys. 42, 134 (1980).

\bibitem {pav}M. Pavsic, Adv. Appl. Clifford Algebras 28, 89 (2018).

\bibitem {haag}R. Haag, Local Quantum Physics, (Springer-Verlag, Berlin1996).

\bibitem {horwitz}L. P. Horwitz and M. Usher, Found. Phys. Lett. 4, 289 (1991).

\bibitem {greiner-rqm}W. Greiner, Relativistic Quantum Mechanics (Springer,
Berlin, 1996).

\bibitem {nitta}H. Nitta and T. Kudo, and H. Minowa, Am. J. Phys. 67, 966 (1999).

\bibitem {our_dirac}M. Alkhateeb, X. Gutierrez de la Cal, M. Pons, D. Sokolovski, and A. Matzkin, Phys. Rev. A 103,
042203 (2021).

\bibitem {schweber}S. S. Schweber, An Introduction to Relativistic Quantum
Field Theory (Dover, New York, 2005).

\bibitem {greiner}W. Greiner, B. Moller, and J. Rafelski, Quantum
Electrodynamics of Strong Fields (Springer, Berlin, 1985), Chaps. 5 and 10.

\bibitem {peskin}M. E. Peskin, D. V. Schroeder, An introduction to quantum
field theory, (Addison-Wesley, Boston, 1995).

\bibitem {our_kg}X. Gutierrez de la Cal, M. Alkhateeb, M. Pons, A. Matzkin and
D. Sokolovski, Sci. Rep. 10, 19225 (2020).

\bibitem {grobe_2023}Gong, C., Li, Y.J., Xi, T.T. et al, Eur. Phys. J. D 77, 4 (2023)
\end{thebibliography}
\end{document}